\documentclass{article}
\usepackage{amsmath, amssymb, amsthm, bbm, hyperref, enumitem, multirow, siunitx, graphicx, float}
\usepackage{doi}
\usepackage{arxiv}
\usepackage[top=3cm, bottom=3cm, right=3cm, left=3cm]{geometry}
\usepackage[round]{natbib}
\usepackage{xcolor,colortbl}  % Coloured text etc.
\definecolor{shaded}{HTML}{E8E8E8}

\usepackage{caption}
\usepackage{xcolor}
\hypersetup{
  colorlinks   = true, 
  urlcolor     = blue, 
  linkcolor    = red, 
  citecolor   = blue
}
\usepackage{color}

\providecommand{\keywords}[1]
{
  \small	
  \textbf{\textit{Keywords---}} #1
}

\newcommand{\reals}{\mathbb{R}}

\newcommand{\prob}{\mathbb{P}}

\newcommand{\given}{\,|\,}

\newcommand{\x}{\mathbf{x}}

\newcommand{\lt}{<}

\newtheorem*{corollary*}{Corollary}

\title{Stop using limiting stimuli as a measure of sensitivities of \\ energetic materials}
\author{Dennis Christensen \and Geir Petter Novik}
\date{\small Norwegian Defence Research Establishment (FFI)}

\defcitealias{stanag4489}{NATO, 1999}
\defcitealias{stanag4487}{2009a}
\defcitealias{stanag4488}{2009b}

\defcitealias{UN2023manual}{United Nations, 2023}
\defcitealias{UN2025model}{2025}

\defcitealias{mil-std-331D}{U.S.~Department of Defense, 2017}
\defcitealias{burke2017binary}{Burke and Truett, 2017}
\defcitealias{baker2021overview}{Baker, 2021}
\defcitealias{emtap2016test43a}{EMTAP, 2016a}
\defcitealias{emtap2016test44a}{2016b}
\defcitealias{btselem2025our}{B'Tselem, 2025}
\defcitealias{clemmensen2024explosive}{Clemmensen, 2024}
\defcitealias{humanity2025explosive}{Humanity \& Inclusion, 2025}

\begin{document}

\maketitle

\begin{abstract}
    Accurately estimating the sensitivity of explosive materials is a potentially life-saving task which requires standardised protocols across nations. One of the most widely applied procedures worldwide is the so-called `1-In-6' test from the United Nations (UN) Manual of Tests in Criteria, which estimates a `limiting stimulus' for a material. In this paper we demonstrate that, despite their popularity, limiting stimuli are not a well-defined notion of sensitivity and do not provide reliable information about a material's susceptibility to ignition. In particular, they do not permit construction of confidence intervals to quantify estimation uncertainty. We show that continued reliance on limiting stimuli through the 1-In-6 test has caused needless confusion in energetic materials research, both in theoretical studies and practical safety applications. To remedy this problem, we consider three well-founded alternative approaches to sensitivity testing to replace limiting stimulus estimation. We compare their performance in an extensive simulation study and apply the best-performing approach to real data, estimating the friction sensitivity of pentaerythritol tetranitrate (PETN).
\end{abstract}


\section{Introduction}\label{sec:introduction}
The assessment of the sensitivity of explosive materials to mechanical stimuli represents a fundamental aspect of energetic material science and safety engineering. Mechanical stimuli such as impact and friction can provide sufficient localised energy to initiate chemical decomposition, which in turn may lead to deflagration or detonation. The degree to which an explosive responds to such stimuli defines its hazard classification and dictates the precautions required during handling, storage and transport. Without systematic sensitivity testing, the risks of accidental initiation would expose personnel and infrastructure to unacceptable hazards. By quantifying sensitivity through standardised methods, it is possible to establish reproducible metrics that inform both regulatory frameworks and material design. Importantly, sensitivity testing is not only relevant for newly produced explosives, as energetic materials can undergo chemical and physical changes over their lifespan that significantly influence their sensitivity and stability. Factors such as moisture ingress, temperature fluctuations, mechanical stress and the natural degradation of binders or stabilisers contribute to the breakdown of explosive compounds, producing more sensitive by-products which alter their characteristics \citep{novik2024increased}. Sensitivity testing is therefore essential for assessing stored explosives, aged ammunition, and explosive remnants of war, of which there already exist millions of tonnes worldwide \citep{novik2022analysis}. With the war of aggression in Ukraine and the genocide in Gaza, clearance of such materials will remain a major hazard for decades to come \citepalias{clemmensen2024explosive, btselem2025our, humanity2025explosive}.

To estimate an energetic material's sensitivity to stimuli like impact or friction, a series of binary trials are conducted. In each trial we stimulate the material at a level $x$ and observe a binary outcome $y\in\{0, 1\}$, namely explosion ($y=1$) or no-explosion ($y=0$). The goal is to estimate the material's sensitivity distribution $F:\reals\to[0, 1]$, a cumulative distribution function (cdf) with the defining property that $F(x) = \prob(y = 1 \given x)$. That is, $F(x)$ is the probability that an explosion will occur when stimulating the material at level $x$. Having conducted $n$ trials, we obtain data $(x_1, y_1), \dots, (x_n, y_n)$, based on which the sensitivity distribution $F$ is estimated.

Any procedure for estimating sensitivity needs to address two key questions, namely (i) how should the stimuli $x_1, \dots, x_n$ be chosen?, and (ii) how should the sensitivity distribution $F$ be estimated? It is too ambitious to try to estimate every single point on the curve $F$ accurately with only a limited number of trials. Therefore, most sensitivity testing procedures focus instead on estimating a single, or a few, key quantities which summarise the behaviour of $F$, like a specific quantile $\xi_{100p} = F^{-1}(p)$ for some $0 < p < 1$, i.e.~the stimulus which causes an explosion with probability $p$. Most, like the up-and-down (Bruceton), Langlie and Neyer procedures, impose a parametric\footnote{In other fields like dose-finding, the up-and-down design is also employed but it is more commonly combined with nonparametric estimation methods like isotonic regression \citep{oron2024up}.} model on $F$ and estimate the median $\xi_{50}$ from the estimated model parameters \citep{dixon1948method, langlie1962reliability, neyer1994optimality}. The first of these is imposed in NATO's standardised agreements \citepalias{stanag4489, stanag4487, stanag4488} and in some United Nations test procedures \citepalias{UN2023manual}. Other designs, like the Robbins--Monro, Robbins--Monro--Joseph and 3pod procedures, are designed so that the input sequence $(x_i)$ itself converges in probability directly to the target quantile \citep{robbins1951stochastic, joseph2004efficient, wu2014three}. With $\xi_{100p}$ being a well-defined and interpretable statistical quantity, the uncertainty of the estimate can also be addressed\footnote{Asymptotic normality of the maximum likelihood estimator has not been formally verified for the Langlie or Neyer procedures.} via confidence intervals \citep{sacks1958asymptotic, tsutakawa1967asymptotic, rosenberger1997asymptotic, christensen2025asymptotic}. This is a particularly important and unfortunately much overlooked point, considering that significant experimental differences in impact sensitivity have been documented across different laboratories \citep{marrs2021sources, mathieu2025misconceptions}.

A different principle altogether from estimating $\xi_{100p}$ is that of estimating a \emph{limiting stimulus} of a material, like the limiting impact energy (for impact sensitivity) or limiting friction load (for friction sensitivity). This is the approach prescribed by the United Nations for its recommended tests \citepalias{UN2023manual, UN2025model}, as well as national standards like the U.S. Department of Defense \citep{dod1998ammunition}, the UK Energetic Materials Testing Assessment Policy Manual of Tests \citepalias{emtap2016test43a, emtap2016test44a} and research institutions like the Lawrence Livermore National Laboratory \citep{hsu2008brief}. For this reason, limiting stimuli are widely employed for measuring sensitivities of energetic materials worldwide \citep{meyer2007explosives, doherty2018making}. In estimating a limiting stimulus, the inputs $x_1, \dots, x_n$ of the binary trials come from a fixed set of values. Starting at some initial value from the set, a binary trial is conducted. If the result is positive (explosion), then the stimulus is decreased to the next lower value. This procedure continues until a stimulus is reached at which $K$ consecutive negative responses (no-explosions) are observed, for some pre-specified number $K$. Depending on convention, either the final or penultimate level of the experiment (i.e.~the first stimulus with no explosions or the last with at least one explosion) is then defined as the limiting stimulus of the material, as a measure of its sensitivity. The estimate is intended to be conservative, which is important to prevent (potentially fatal) accidents when handling and transporting explosives.

In the present paper we argue that, in spite of their wide use, limiting stimuli should be abandoned as a measure of sensitivities of energetic materials. Our justification for this position is that a limiting stimulus is not a well-defined statistical quantity. This is in stark contrast with for instance quantiles $\xi_{100p}$, which are intrinsic to the material's sensitivity distribution. This means that any procedure for estimating say $\xi_{50}$ (like up-and-down, Langlie, Neyer, etc.) are effectively targeting the same quantity, and, with enough data, the estimates of these procedures, along with their confidence intervals, will coincide. When estimating a limiting stimulus, on the other hand, a change in the parameters of the test procedure will actually change the value of the limiting stimulus of the material. This directly contradicts the claim in Section 1.1.5 of the UN manual, which asserts that, apart from physical parameters, ``the outcome of the tests in this Manual is predominantly related to the intrinsic properties of the substance being tested.''

Whilst previous studies have noted the ill-defined nature of limiting stimuli \citep{collet2021review, esposito2024review}, this paper aims to demonstrate their detrimental impact on energetic materials research. In particular, we will use examples from the literature to show how limiting stimuli are continually misinterpreted and employed in studies where quantities like $\xi_{50}$ (or another suitable quantile) would have been far more appropriate. We do not seek to blame any individual researcher; the whole field of energetic materials research is at fault, and, consequently, collectively suffers from this confusion. In fact, one of our examples is taken from a study by the first author of the present paper, who was unaware of the ambiguity surrounding limiting stimuli at the time. We shall outline how a more robust statistical analysis of sensitivity---and in particular, the consideration of uncertainty via confidence intervals---could better substantiate the results of the studies considered.

At this point it should be mentioned that the UN manual uses limiting stimuli as an intermediate step in a binary classification of a material as sensitive or insensitive. For instance, a material is considered sensitive to friction if its limiting frictional load is less than \SI{80}{\N}, and insensitive otherwise \citepalias[Section 13.5.1.4]{UN2023manual}. In her thorough review, \citet{doherty2018making} also notes that limiting stimuli should only be used as a tool for binary classification. One might therefore argue that a warning against the use of limiting stimuli as a measure of sensitivity is superfluous, since this is not necessarily their intended use. However, this distinction is commonly ignored, and there are myriad examples in the energetic materials literature where limiting stimuli are being used as a standalone measure of sensitivity, both in industry and in research. Moreover, if one wishes to construct a binary classification scheme for explosives, it too ought to rely on a sound statistical foundation, so that the classification criterion is well-founded and not dependent on arbitrary experimental parameters. This would also enable the operator to assess the uncertainty of the classification, which the UN manual does not. 

As alternatives to estimating limiting stimuli, we examine two other sensitivity estimation procedures: the biased coin design (BCD) by \citet{durham1993convergence} and the Robbins--Monro--Joseph procedure (RMJ) by \citet{joseph2004efficient}. Like the intended goal of measuring limiting stimuli, these procedures enable the estimation of quantiles more conservative than the median $\xi_{50}$, whilst also being well-defined statistical procedures with quantifiable uncertainty via confidence intervals. We compare their performance against the UN manual procedure, both via simulations and real-world experimental data measuring the friction sensitivity of pentaerythritol tetranitrate (PETN).

\section{Limiting stimuli}
In this section we go through the procedures for estimating limiting stimuli from the UN manual of tests and criteria \citepalias{UN2023manual}, and demonstrate their shortcomings.

\subsection{Description of tests}\label{sec:tests}
The manual (Section 13:~Test Series~3) describes a total of eleven test procedures, seven for impact and four for friction. Of the former category, three tests target a limiting impact energy, named the BAM\footnote{Short for Bundesanstalt f\"{u}r Materialforschung.} fallhammer (the recommended test), the \SI{30}{\kg} fallhammer test and the Impact sensitivity test. We label these tests as I1, I2 and I3, respectively. For friction sensitivity, there are two test methods estimating a limiting frictional load: the BAM friction apparatus (the recommended test) and the Friction sensitivity test, labelled here as F1 and F2, respectively.

Before discussing the statistical aspects of the above procedures, we will briefly explain the physical setup of the two recommended tests for impact and friction, i.e.~I1 and F1. The former employs the BAM fallhammer, a test apparatus consisting of a vertically aligned drop weight which is released from a predetermined height onto a sample of material confined between two steel anvils. Here, a sample of a defined volume (usually \SI{40}{\milli\metre\cubed}) is enclosed in a metal cylinder, and the test is conducted under controlled environmental conditions. When the weight impacts the sample, a positive reaction such as detonation is typically indicated by sound, flame or smoke. Test F1 uses the BAM Friction apparatus, which assesses the friction sensitivity of an energetic material by subjecting it to a defined normal force and lateral movement. The test setup includes a porcelain plate and a pivoted arm with a porcelain peg that presses against the sample. The arm is loaded with a specific force (weight) placed at one of the predefined locations on the pivotal arm (notch), and the peg is moved a fixed distance across the sample surface at constant speed. The frictional loads obtained by placing the weights in the different notches are given in Table~\ref{tab:frictional-loads}. A positive result in this test may manifest as ignition, explosion, or other visible signs of decomposition.

\begin{table}[t]
\centering
\caption{Frictional loads (in Newton) used for the BAM friction apparatus (F1).} \label{tab:frictional-loads}
    \begin{tabular}{crrrrrrr}
        \hline \multirow{2}{*}{\vspace{-1ex}\shortstack{Weight \\ No.}} & \multicolumn{1}{c}{\multirow{2}{*}{Mass}} & \multicolumn{6}{c}{Notch No.} \\
        && 1 & 2 & 3 & 4 & 5 & 6 \\
        \hline
        B1 & \SI{0.28}{\kg} & 5 & 6 & 7 & 8 & 9 & 10 \\
        B2 & \SI{0.56}{\kg} & 10 & 12 & 14 & 16 & 18 & 20 \\
        B3 & \SI{1.12}{\kg} & 20 & 24 & 28 & 32 & 36 & 40 \\
        B4 & \SI{1.68}{\kg}& 30 & 36 & 42 & 48 & 54 & 60 \\
        B5 & \SI{2.24}{\kg} & 40 & 48 & 56 & 64 & 72 & 80 \\
        B6 & \SI{3.36}{\kg} & 60 & 72 & 84 & 96 & 108 & 120 \\
        B7 & \SI{4.48}{\kg} & 80 & 96 & 112 & 128 & 144 & 160 \\
        B8 & \SI{6.72}{\kg} & 120 & 144 & 168 & 192 & 216 & 240 \\
        B9 & \SI{10.08}{\kg} & 180 & 216 & 252 & 288 & 324 & 360 \\
        \hline
    \end{tabular}
\end{table}

We now turn to statistical aspects of the five given tests. They all follow the same general principle outlined in Section~\ref{sec:introduction}, namely that binary trials are conducted at a fixed set $\mathcal{S}$ of stimuli in decreasing order and terminate when a level is reached at which $K$ consecutive negative responses are observed, for some number $K$. Beyond this commonality, however, the tests differ profoundly in how they are defined. Firstly, the two recommended tests (I1 and F1) use $K=6$, meaning the experiment terminates after observing six consecutive trials with negative responses. Tests I3 and F2, on the other hand, use $K=25$, whilst I2 uses $K=3$. Also, I1 and F1 define the limiting stimulus as the \emph{smallest} stimulus which yields at least one positive response (i.e.~the penultimate stimulus level of the experiment), whereas the other tests define it as the \emph{largest} stimulus which yielded the $K$ negative responses (i.e.~the stimulus level at which the test terminates). We shall refer to these two different definitions of a limiting stimulus as type I and II, respectively.

In I2 and F1, the first trial is conducted at the maximum value of $\mathcal{S}$. I1 and I3, on the contrary, contain an additional initial stage, which begins with a binary trial at an initial value roughly in the middle of the range of the set $\mathcal{S}$. If the result of this trial is positive, the test continues like the others, with trials at the next lowest value. If, however, it is positive, then the next trial is conducted at the next higher value of $\mathcal{S}$, and this process continues until a positive response is observed, upon which the procedure continues as normal. Finally, F2 does not mention whether an initial stage is to be included or not, leaving its initial stimulus undefined.

Three of the tests, namely I2, I3 and F2, have a fixed interval between all the values in the set $\mathcal{S}$, whereas I1 lists the elements of $\mathcal{S}$ explicitly: ``The \SI{1}{\kg} drop weight is used at fall heights of 10, 20, 30, 40 and \SI{50}{\cm} (impact energy 1 to \SI{5}{\J}); the \SI{5}{\kg} drop weight for fall heights of 15, 20, 30, 40, 50 and \SI{60}{\cm} (impact energy 7.5 to \SI{30}{\J}) and the \SI{10}{\kg} drop weight for fall heights of 35, 40 and \SI{50}{\cm} (impact energy 35 to \SI{50}{\J}).'' Finally, F1 states that ``The use of different weights in notches results in loads on the peg of 5 - 10 - 20 - 40 - 60 - 80 - 120 - 240 - \SI{360}{\N}. If necessary, intermediate loads may be used.'' This leaves the set $\mathcal{S}$ undefined, since it is up to the individual operator to define when the use of intermediate loads is `necessary'. This ambiguity may seem like a minor issue, but as we shall see shortly, this has a major effect on the results of the test. Looking at Table~\ref{tab:frictional-loads}, we see that there are multiple ways to include intermediate loads. For instance, a particularly diligent operator might always choose to include as many intermediate loads as possible, yielding the sequence 360, 324, 288, 252, 240, etc. Other interpretations are also possible. This ambiguity is particularly alarming, as F1 is the manual's recommended test for friction sensitivity testing.

Table~\ref{tab:UN-manual-tests} summarises all five tests and highlights their differences. There are indeed many discrepancies, and both friction tests (including the recommended test) contain undefined elements. It may be tempting to think that unifying the entries of Table~\ref{tab:UN-manual-tests} would resolve all problems with limiting stimuli. Whilst that would be an improvement, it would not address the root problem: the limiting stimulus is not a well‑defined statistical phenomenon, and its definition inevitably depends on arbitrary experimental parameters. Thus the divergent entries in Table~\ref{tab:UN-manual-tests} should be treated as a symptom, not the root cause, of the issue.

\begin{table}
\centering
    \caption{Variants of the procedures from the UN manual.} \label{tab:UN-manual-tests}
    \begin{tabular}{cccccccc}
    \hline Sensitivity & Section & ID & Name & $K$ & Initial stage & Step size & Type \\
    \rowcolor{shaded} & 13.4.2 & I1 & BAM Fallhammer & 6 & Yes & Varying & I \\
    \rowcolor{shaded} Impact & 13.4.4 & I2 & \SI{30}{\kg} Fallhammer test & 3 & No & Fixed & II \\
    \rowcolor{shaded}& 13.4.6 & I3 & Impact sensitivity test & 25 & Yes & Fixed & II \\
    \multirow{2}{*}{Friction} & 13.5.1 & F1 & BAM friction apparatus & 6 & No & Undefined & I \\
    & 13.5.3 & F2 & Friction sensitivity test & 25 & Undefined & Fixed &  II \\
    \hline
    \end{tabular}
\end{table}

\subsection{Changing test parameters}\label{sec:changing-test-parameters}
\begin{figure}
    \begin{center}
        \includegraphics[scale=.55]{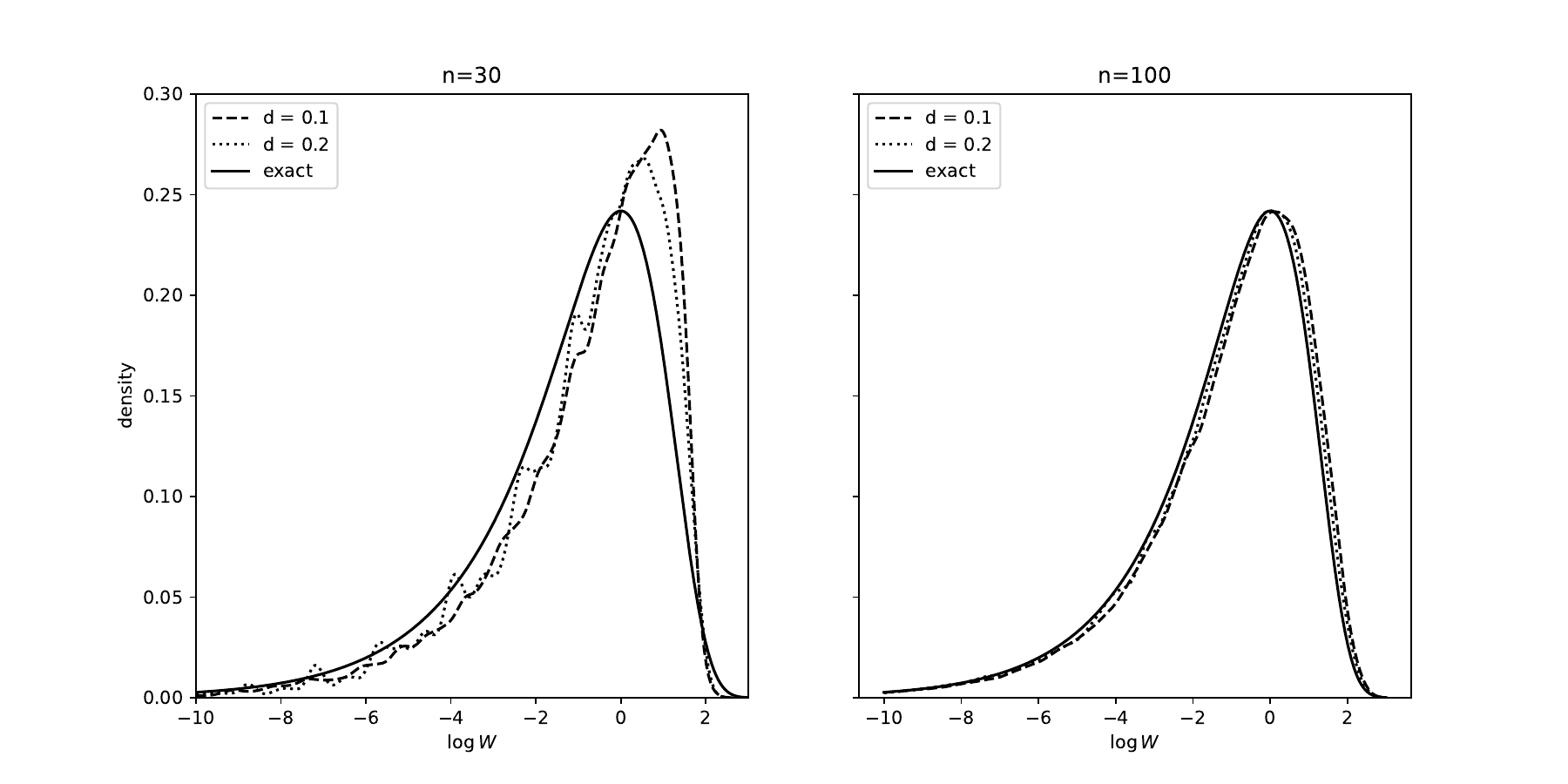}
    \end{center}
    \caption{Kernel density estimates $\log W$ for $n=30, 100$ and $d=0.1, 0.2$, based on $S=100{,}000$ datasets. The solid curve shows the exact distribution of the logarithm of a $\chi^2_1$ random variable.}
    \label{fig:changing-parameters1}
\end{figure}

To highlight the effect of arbitrary parameters on limiting stimuli, we consider a simple simulation study in which we measure the friction sensitivity of an imagined energetic material whose sensitivity is given by
    \begin{equation}\label{eq:probit_model}
        \prob(Y = 1\given X=x) = \Phi(\theta^\top z),
    \end{equation}
where $\theta = (\alpha, \beta)^\top$, $z = (1, \log x)^\top$ and $\Phi$ denotes the standard normal cumulative distribution function (cdf). As ground truth, we use $\theta_0 = (-9.1258, 2.0473)^\top$. These values are the maximum likelihood estimates of $\alpha$ and $\beta$ based on the data from \citet{novik2025characteristics}, using model~\eqref{eq:probit_model}. We will compare the performance of test F1 from the UN manual with the up-and-down procedure, which begins with an initial binary trial at some pre-specified input $x_1$ and then subsequently sets
\begin{equation}\label{eq:up-and-down}
    \log x_{i+1} = \begin{cases}\log x_i - d &\text{if}\;y_i = 1, \\ \log x_i + d &\text{if}\;y_i=0,\end{cases}
\end{equation}
for some pre-specified parameter $d > 0$, called the step size. Using model~\eqref{eq:probit_model}, the observed Fisher information matrix takes the form
    \begin{equation}\label{eq:fisher-information}
        J_n = \sum_{i=1}^n\frac{\phi(\theta^\top z_i)}{\Phi(\theta^\top z_i)\{1 - \Phi(\theta^\top z_i)\}}zz^\top,
    \end{equation}
where $\phi$ is the standard normal probability density function. Due to the dependence introduced by the adaptive procedure~\eqref{eq:up-and-down}, asymptotic normality of the maximum likelihood estimator $\widehat\theta$ for model~\eqref{eq:probit_model} is not automatic, but various authors have established via martingale theory that $\sqrt{n}(\widehat\theta - \theta_0)$ converges to a zero-mean normally distributed variable with covariance matrix $V = J^{-1}$, where $J$ is the probability limit of $J_n$ as $n\to\infty$. Writing $f_0 = (1, -\alpha_0 / \beta_0)^\top$, this implies that $W = (f_0^\top \widehat\theta)^2/\{f_0^\top V f_0\}$ is asymptotically chi-squared distributed with one degree of freedom. This is the basis of confidence intervals for $\log F_{50} = -\alpha / \beta$ via Fieller's theorem \citep{fieller1954some}. In our simulations, we generated $S=100{,}000$ data sets $\{(x_i, y_i)\}_{i=1}^n$ using $n=30$ and $n=100$ using the up-and-down method with initial frictional load $x_1 = \SI{360}{\N}$ and step sizes $d_1 = 0.1$ and $d_2 = 0.2$. For each dataset\footnote{For $n=30$, the maximum likelihood estimates were well-defined for more than 99\% of the datasets. For $n=100$ they were well-defined for all datasets.} we computed the variable $\log W$, resulting in the empirical distributions in Figure~\ref{fig:changing-parameters1}. This illustrates the convergence of $W$ to the chi-squared distribution, and even with a sample size of $n=30$, we get decent results, and previous simulations have demonstrated that confidence intervals for $\log F_{50}$ via Fieller's theorem yield satisfactory coverage probabilities with $n=30$ \citep{christensen2023improved}. The key takeaway however is that the convergence holds regardless of the value of the step size $d$. Whilst certain values of $d$ lead to more rapid convergence than others, the target $\log F_{50}$ does not change as we change experimental parameters like $d$.

\begin{figure}
    \begin{center}
        \includegraphics[scale=.7]{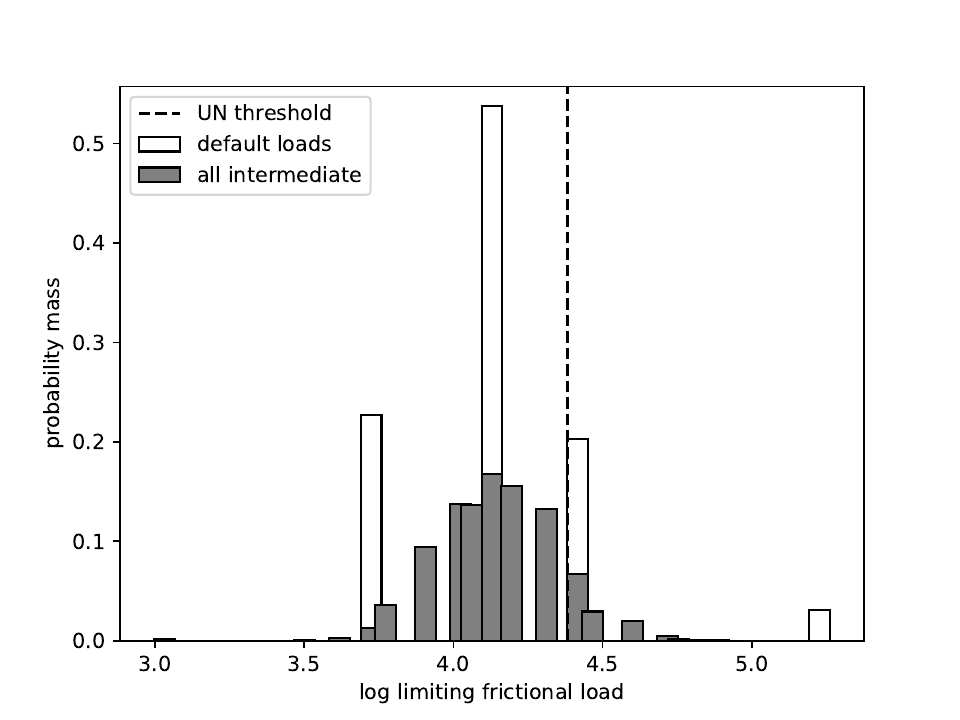}
    \end{center}
    \caption{Limiting frictional stimuli (log scale) based on $S=100,000$ simulations using two different interpretations of the loads available for test F1 from the UN manual. The dashed vertical line gives the manual's threshold for sensitive versus insensitive material.}
    \label{fig:changing-parameters2}
\end{figure}

\begin{figure}
    \begin{center}
        \includegraphics[scale=.7]{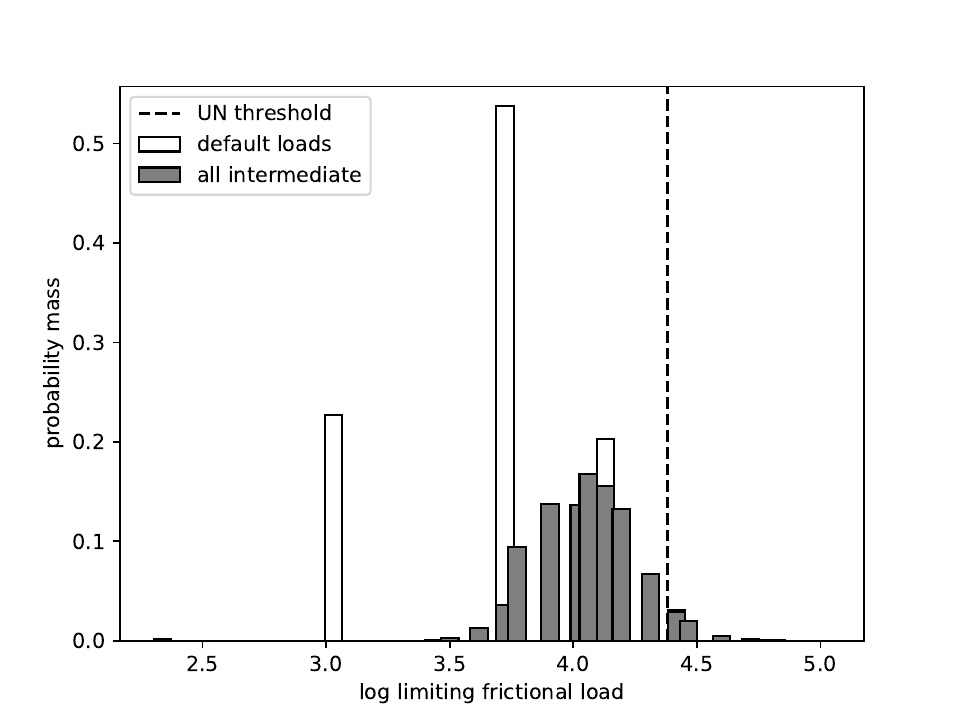}
    \end{center}
    \caption{Limiting frictional stimuli (log scale) based on $S=100,000$ simulations using two different interpretations of the loads available for test F1 from the UN manual, using type II tests rather than type I. The dashed vertical line gives the manual's threshold for sensitive versus insensitive material.}
    \label{fig:changing-parameters3}
\end{figure}

Having illustrated the robustness of the up-and-down procedure, let us now turn to F1, the UN manual's recommended test for friction sensitivity. In light of the discussion pertaining to Table~\ref{tab:UN-manual-tests}, we use simulations to examine the effect of different interpretations of the available frictional loads. First, we generated $S=100,000$ limiting frictional loads using the default loads provided in the manual, i.e.~keeping notch 6 fixed and only varying the weight (apart from the smallest load of \SI{5}{\N}, for which one must use notch 1). Next, we repeated the experiment but using all intermediate loads available. The distributions of the limiting frictional loads are plotted in Figure~\ref{fig:changing-parameters2}. Unlike the densities in Figure~\ref{fig:changing-parameters1}, these distributions are discrete due to the nature of the test procedure. We see a large difference between the two distributions, particularly in terms of their dispersion. The two distributions are indeed centred around the same point, but this turns out to be a coincidence rather than a general property of the test procedures. Indeed, if we repeat the simulations but change the definition of the test procedures from type I to type II (as defined in Table~\ref{tab:UN-manual-tests}, we obtain the results in Figure~\ref{fig:changing-parameters3}. Similar changes in the results can be obtained by altering other experimental parameters, like the set $\mathcal{S}$ of available values. This shows that the limiting frictional load depends heavily on the experimental parameters of the test, illustrating that the limiting stimulus is not a well-defined statistical phenomenon. As mentioned in the Introduction, the UN manual uses limiting stimuli for a binary classification between sensitive versus insensitive compounds. In Figure~\ref{fig:changing-parameters2}, we see the threshold used in test F1, namely \SI{80}{\N} (on a log scale). The simulations show that the two different interpretations would lead to the material being classified as sensitive 76.6\% and 87.6\% of the time, respectively. Hence, different choices in experimental parameters result in large classification probabilities for a single compound. Additionally, any confidence interval for the limiting stimulus or the binary classification of the material would depend on the test parameters as well, essentially implying that it is not possible to evaluate the uncertainty of the estimates obtained.

\defcitealias{stanag4487}{NATO, 2009} % need to redefine this alias here for the citation to look right
On average, it took 15.12 and 36.98 trials for the F1 tests to terminate, respectively. In particular, when using all intermediate loads, we need on average more trials than that required by the up-and-down procedure from NATOs standardised agreement on friction sensitivity, which is 30 \citepalias{stanag4487}. In other words, measuring limiting stimuli should not necessarily be seen as a `quick and dirty' way to obtain a rough estimate of a material's sensitivity. As this study indicates, it is possible to obtain far more statistical information with roughly the same, or only a few more trials, using a well-founded experimental design and estimation procedure.

\subsection{A real data example}\label{sec:PETN1}
In addition to the simulation study, we also employed the two interpretations of the UN manual in real experiments with PETN. The sample used in the analyses was `PETN Wax NSP452' grains with 7.7\% content of wax, produced by EURENCO Bofors in Karlskoga, Sweden, released for sale following Certification of Compliance/Analysis on November 29th 2018. The results of the test using the default frictional loads (i.e.~keeping notch 6 fixed) are reported in Table~\ref{tab:one-of-six-petn1}. As \SI{80}{\N} was the last level with at least one positive trial, this is the resulting limiting frictional load, and the material is classified as insensitive. Table~\ref{tab:one-of-six-petn2} contains the corresponding results from the test using all intermediate loads available. With \SI{48}{\N} as the last level with a positive response, this is the limiting frictional load, resulting in a classification as sensitive. Naturally, any diagnostic test based on statistical modelling will contain misclassifications, but the main disadvantage here is that the classification probabilities depend on non-intrinsic factors of the test itself, and we have no quantification of the probability of achieving a correct classification.

\begin{table}
    \centering
    \caption{Results of friction sensitivity testing of PETN using test F1 from the UN manual with the default loads provided in the test.}\label{tab:one-of-six-petn1}
    \begin{tabular}{lcccccc}
        \hline
        Trial \# & 1 & 2 & 3 & 4 & 5--6 & 7--12 \\
        Weight & B9 & B8 & B7 & B6 & B5 & B4 \\
        Notch & 6 & 6 & 6 & 6 & 6 & 6 \\
        Load & 360 & 240 & 160 & 120 & 80 & 60 \\
        Result(s) & 1 & 1 & 1 & 1 & 01 & 000000 \\
        \hline
    \end{tabular}
\end{table}

\begin{table}
\centering
    \caption{Results of friction sensitivity testing of PETN using test F1 from the UN manual with all intermediate loads available.}\label{tab:one-of-six-petn2}
    \begin{tabular}{lcccccccccccccc}
    \hline
        Trial \# & 1 & 2 & 3 & 4 & 5 & 6 & 7 & 8 & 9 & 10 & 11 & 12 & 13 & 14 \\
        Weight & B9 & B9 & B9 & B9 & B8 & B8 & B8 & B9 & B8 & B7 & B7 & B7 & B6 & B7 \\
        Notch & 6 & 5 & 4 & 3 & 6 & 5 & 4 & 1 & 3 & 6 & 5 & 4 & 6 & 3 \\
        Load & 360 & 324 & 288 & 252 & 240 & 216 & 192 & 180 & 168 & 160 & 144 & 128 & 120 & 112 \\
        Result & 1 & 1 & 1 & 1 & 1 & 1 & 1 & 1 & 1 & 1 & 1 & 1 & 1 & 1 \\
        \hline \\
    \end{tabular}

    \begin{tabular}{ccccccccccc}
\hline 15 & 16 & 17 & 18 & 19 & 20 & 21--22 & 23 & 24--26 & 27--31 & 32--37 \\
B6 & B6 & B6 & B5 & B5 & B5 & B4 & B5 & B4 & B4 & B4 \\
5 & 4 & 3 & 6 & 5 & 4 & 6 & 3 & 5 & 4 & 3 \\
108 & 96 & 84 & 80 & 72 & 64 & 60 & 56 & 54 & 48 & 42 \\
1 & 1 & 1 & 1 & 1 & 1 & 01 & 1 & 001 & 00001 & 000000 \\
\hline
\end{tabular}
\end{table}

\section{Consequences}
Having demonstrated the ambiguous nature of limiting stimuli, we will now examine some examples from the literature which illustrate the adversarial effect this notion of sensitivity has had on energetic materials research. We will also discuss how these shortcomings could be overcome by a more rigorous statistical treatment of sensitivity testing.

A common mistake made in the field is to treat limiting stimuli and $\xi_{50}$ as equivalent notions of sensitivity. This is particularly prevalent in correlation studies aiming to predict the sensitivity of materials based on their chemical properties via statistical or machine learning models. The first author of the present paper is also guilty of this fault in an earlier study, namely \citet{jensen2020models}. Here, linear models for predicting impact sensitivity are compared, where the best-performing alternative uses a single covariate: the bond dissociation energy divided by the explosion temperature. The data for nitroaromatics and nitramines are taken from \citet{wilson1990explosive, storm1990sensitivity}, who use $\xi_{50}$, but the data for nitrate esters (solid and liquid) are taken from \citet{meyer2007explosives}, who use limiting impact energies. Since the limiting stimulus itself depends on arbitrary test parameters (cf.~Section~\ref{sec:changing-test-parameters}), so do the regression parameters. This deprives the estimated model of its interpretable properties, hindering the possibility of gaining new chemical insights from data, which is often the goal of fitting such regression models in the first place. One of the main challenges of linking the sensitivity of energetic materials to their chemical properties is precisely that there are many factors which are known to affect sensitivity (like heat of detonation, detonation velocity, oxygen balance, crystal morphology, etc.), and many sources of statistical variation (operator-related judgement of explosion versus no-explosion, type of testing apparatus, etc.). Using limiting stimuli rather than a well-defined notion of sensitivity, like $\xi_{50}$, only introduces additional noise, and is advised against, particularly in situations with limited data, like that of \citet{keshavarz2015new}, where there is danger of overfitting. Furthermore, predictive models for sensitivity based on machine learning are sometimes trained on large amalgamated datasets comprising both $\xi_{50}$ measurements and limiting stimuli \citep{marrs2023chemical}, which may introduce additional noise and obfuscate the regression task.

In addition to treating $\xi_{50}$ and limiting stimuli as equivalent notions of sensitivity, another prevalent misconception in energetic materials research is to associate the limiting stimulus with the quantile $\xi_{100/K}$ [a non-exhaustive list includes \citet{chapman1998studies, walley2006crystal, alouaaaamari2008statistical, keicher2009lab, gruhne2020OZM, jeunieau2024phlegmatization}]. This most likely stems from the use of the term `1-In-$K$' to denote the tests from the UN manual (see Section~\ref{sec:tests}). For instance, the UK Energetic Materials Testing Assessment Policy Manual of Tests (EMTAP) refers to both I1 and F1 as the `1-In-6' tests \citepalias{emtap2016test43a, emtap2016test44a}. In this regard, the erroneous interpretation of a limiting stimulus as the one-sixth quantile, or $\xi_{16.7}$, is a very understandable one. However, although the tests indeed terminate at the stimulus where at most one of six trials are positive, this does not imply that the test estimates the one-sixth quantile. Indeed, the simulations from Section~\ref{sec:changing-test-parameters} clearly demonstrate that limiting stimuli cannot be interpreted as an estimate of any specific quantile of the sensitivity distribution, and that it can cover a wide range of quantiles, depending on the external parameters of the test. This is important, as advocates of measuring limiting stimuli over $\xi_{50}$ argue that it yields a more conservative notion of sensitivity. It is true that in practice, using the UN manual will result in a lower value than if one measures $\xi_{50}$, but the degree of conservativeness depends on external test parameters and is outside the control of the operator. In particular, it may not resemble $\xi_{16.7}$ at all. In conclusion, the wrongful interpretation of a limiting stimulus as $\xi_{16.7}$ provides a false sense of confidence in an estimate which in reality is much more ambiguously defined than such an interpretation would suggest. This is particularly regrettable for safety applications, where the interpretation of probabilities is directly linked to risk assessment.

Some studies attempt to establish relationships between limiting stimuli and other notions of sensitivity, like that of \citet{wharton1997relationship}, whose main result also appears in \citet{lefrancois2022characterizing}. Here, linear regression is used to relate limiting to rotary friction sensitivities. However, the parameters of such models are also affected by the external parameters of the tests from the UN manual, just like in models for predicting limiting stimuli from chemical descriptors. We therefore advise against employing these results, as they are unlikely to generalise beyond the set of compounds on which the model is trained.

We conclude this section with a brief discussion about hypothesis testing, an issue which, much like the construction of confidence intervals, has been overlooked in the energetic materials literature. There are many instances in which researchers have two sensitivity datasets $\{(x_i^A, y_i^B)\}_{i=1}^{n^A}$ and $\{(X_i^B, y_i^B)\}_{i=1}^{n^B}$ and would like to assert that the sensitivity distributions which generated these are truly different. The most recent study of those listed above, namely \citet{jeunieau2024phlegmatization}, is a perfect example of this. This important work addresses the disposal of triacetone triperoxide (TATP), a particularly sensitive explosive employed in numerous terrorist attacks. Conventionally, explosive ordnance disposal teams have added diesel oil to make TATP less sensitive during disposal. The authors demonstrate that vacuum oil works better, making the explosive less sensitive to both friction and impact. Unfortunately, since the authors use the sensitivity tests from the UN manual, the difference in sensitivity is examined simply by comparing two different stand-alone numbers (e.g.~\SI{2}{\N} versus \SI{5}{\N} for friction sensitivity). No hypothesis test is conducted. 

In some situations, we are interested in asserting that the underlying sensitivity distributions generating the two datasets are in fact equal. One example is round robin testing across different laboratories to ensure that different research institutions get similar results when estimating the sensitivity of the same material under the same experimental conditions. \citet{krabbendam2016overview} conducted such a test with an impressive number of institutions from different nations to provide results for comparison. Unfortunately, many of the participants insisted on using the test procedures from the UN manual, which means a proper statistical comparison of their results is not possible. In summary, further work on hypothesis testing for sensitivity measurements is highly encouraged, as well as the development of practical guidelines for their implementation.

\section{Alternatives}
We now cover three alternatives to using limiting stimuli as a notion of sensitivity. All three are in principle able to estimate conservative quantiles and have known theoretical properties. Having introduced the procedures, we compare their performance in a simulation study focusing both on estimation accuracy and on confidence intervals. We also present the result of an experiment in which we applied the best working method to measure the friction sensitivity of PETN.

\subsection{The up-and-down design with parametric estimation}
The first alternative we shall consider is simply the standard up-and-down design~\eqref{eq:up-and-down} combined with maximum likelihood estimation using the probit model on a log scale. For confidence intervals, we use Fieller's theorem as in \citet{christensen2023improved}. As this is a purely parametric estimation procedure, we do not expect it to perform well if the model is severely misspecified.

\subsection{The biased coin design}
The second alternative we shall consider is the biased coin design (BCD) by \citet{durham1993convergence}. This is very similar to the classical up-and-down design, but with an extra `coin flip' stage added, where the bias of the coin allows the limiting distribution of the process to target other quantiles than the median $\xi_{50}$ of the sensitivity distribution. Say we aim to estimate $\xi_{100p}$, where $0 < p < 1/2$. After the initial trial at stimulus $x_1$, the updating rule goes as follows. If the previous trial had a positive response, then we always escalate one step like the original up-and-down design. However, if the response was negative, then we flip a coin with probability $p/(1 - p)$ of landing heads. We de-escalate if the coin lands heads, and we stay where we are otherwise. \citet{durham1993convergence} proved that the limiting distribution of the design is centred at the target quantile $\xi_{100p}$. Furthermore, the positive probability of staying still removes the 2-periodicity from the up-and-down design and ensures faster mixing to the equilibrium distribution. If $p > 1/2$, the updating rule is reversed accordingly, see \citet{durham1993convergence} for details.

Whilst it is possible to combine the BCD with maximum likelihood estimation, we use centred isotonic regression (CIR) here for estimation and the creation of confidence intervals, as recommended by \citet{oron2017centered, flournoy2020bias, oron2022understanding}. CIR is an improvement of the original isotonic regression algorithm for binary response data by \citet{ayer1955empirical}, which maximises the log-likelihood
    $$\sum_{i=1}^n y_i\log F(x_i) + (1 - y_i)\log\{1 - F(x_i)\},$$
    over the space of all monotonic functions $F:\reals\to[0, 1]$. To implement the BCD and CIR in practice, we use the R packages \texttt{upndown} and \texttt{cir} \citep{oron2025cir, oron2025upndown}. In particular, we use the recommended method based on local inversion and the Delta method for constructing confidence intervals, as implemented by the function \texttt{udest} in the R package \texttt{upndown} \citep{oron2017centered}.

\subsection{The Robbins--Monro--Joseph procedure}
The second alternative we shall consider is the Robbins--Monro--Joseph procedure (RMJ), a modified version of the classic procedure by \citet{robbins1951stochastic} due to \citet{joseph2004efficient}. For binary trials with inputs on a log scale, the original Robbins--Monro procedure uses the adaptive rule $\log x_{i+1} = \log x_i - a_i (y_i - p)$, where $(a_i)$ is a fixed positive sequence, and $0 \lt p \lt 1$ is fixed. \citet{robbins1951stochastic} showed that $x_i \to \xi_{100p}$ in probability provided $\sum_{i=1}^\infty a_i = \infty$ and $\sum_{i=1}^\infty a_i^2 \lt \infty$, which was strengthened to almost sure convergence by \citet{blum1954approximation}. Thus, the Robbins--Monro procedure yields $x_{n+1}$, the next hypothetical input after $n$ trials, as a nonparametric estimator for $\xi_{100p}$. Furthermore, \citet{chung1954on} established asymptotic normality of the process, enabling the construction of confidence intervals for $\xi_{100p}$. Although the original procedure works well for quantiles near the median, \citet{joseph2004efficient} noted that this is not the case for extreme quantiles, like $p=0.1$ or $p=0.9$. His variant RMJ uses an adaptive rule of the form $\log x_{i+1} = \log x_i - a_i(y_i - b_i)$ instead, where the sequence $(b_i)$ is optimised subject to a suitable criterion, specifically for binary trials. The optimal choices of $a_i$ and $b_i$ have to be approximated, since they depend on the target quantile, which is unavailable to the experimenter. The resulting updating rule is given in \citet{joseph2004efficient} and we do not repeat it here. The recipe for constructing confidence intervals is also provided. It is also worth mentioning that further alterations of this process have been proposed, such as that of \citet{wang2015skewed}, which seem to work better for even more extreme quantiles outside the interval $[0.1, 0.9]$. In this work, we only focus on the RMJ process.

\subsection{Simulations}
We compared the performance of the three above approaches in a simulation study, the results of which are covered here. Replication code can be found in the online supplementary material. We employed the same six distributions as in \citet{joseph2004efficient} as models, namely (a) normal; (b) uniform; (c) logistic; (d) extreme value; (e) skewed logistic; (f) Cauchy. For each model, we investigated how well each approach can estimate and create a 90\% confidence interval for the quantile $\xi_{100p}$, with $p\in\{0.10, 0.25, 0.50, 0.75, 0.90\}$ after $n=30$ and $n=100$ observations, respectively. So for each model choice, quantile $p$ and sample size $n$, we generated $S=10{\,}000$ datasets using the standard up-and-down, BCD and RMJ designs, and estimated $\xi_{100p}$ and constructed a 90\% confidence interval.

The parameters of the approaches were chosen as follows. For the up-and-down design and the BCD, we set $x_1 = \xi_{100p}$ and $d=0.5$. For the RMJ procedure, we use mostly the same hyperparameters as in \citet{joseph2004efficient}, namely initial uncertainty $\tau_1 = 1$ and $x_1$ normally distributed with zero-mean and variance $\tau_1^2$. A key difference, however, is that we do not pretend to have access to the derivative of the true distribution function at the target quantile. That is, in the notation of \citet{joseph2004efficient}, we do not claim to know $\dot{M}(0)$, as this would give the RMJ procedure a significantly unfair advantage over the two other methods, which do not have access to such information. Thus, instead of using $\dot{M}(0)$, we always use $\phi(\Phi^{-1}(p))$, meaning we employ a normal approximation of the derivative of the true distribution function. This is in fact much in line with the other approximations made in \citet{joseph2004efficient}.

\begin{figure}
    \begin{center}
        \includegraphics[scale=.52]{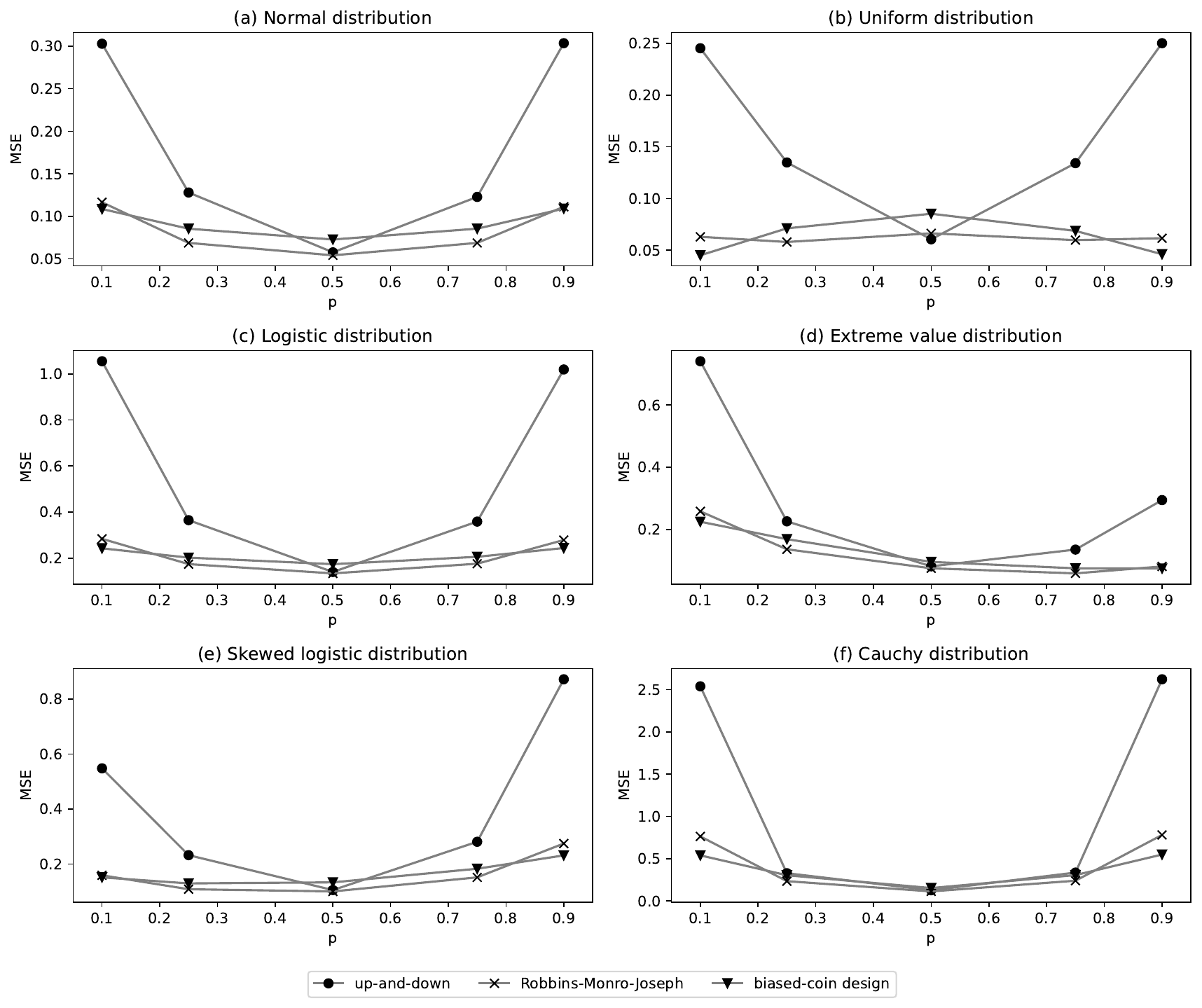}
    \end{center}
    \caption{Mean squared errors (MSEs) from the simulation study after $n=30$ trials.}
    \label{fig:MSE}
\end{figure}

First we compare the estimation abilities of the three approaches. In Figure~\ref{fig:MSE}, we see the mean squared errors (MSEs) they achieved for the target quantiles $\xi_{100p}$ after $n=30$ trials. We clearly see that on the whole, the standard up-and-down design equipped with maximum likelihood estimation performs much worse than both the BCD and the RMJ. Note that this is also true for the normal distribution, where the former approach is actually correctly specified. There does not seem to be any significant difference between the performance of the BCD and RMJ; both approaches yield satisfactory results. We also see that all three methods do better for quantiles close to the median, and, with few exceptions, worse for extreme quantiles. The results for $n=100$ are qualitatively very similar, and can be found in the Appendix.

\begin{figure}
    \begin{center}
        \includegraphics[scale=.52]{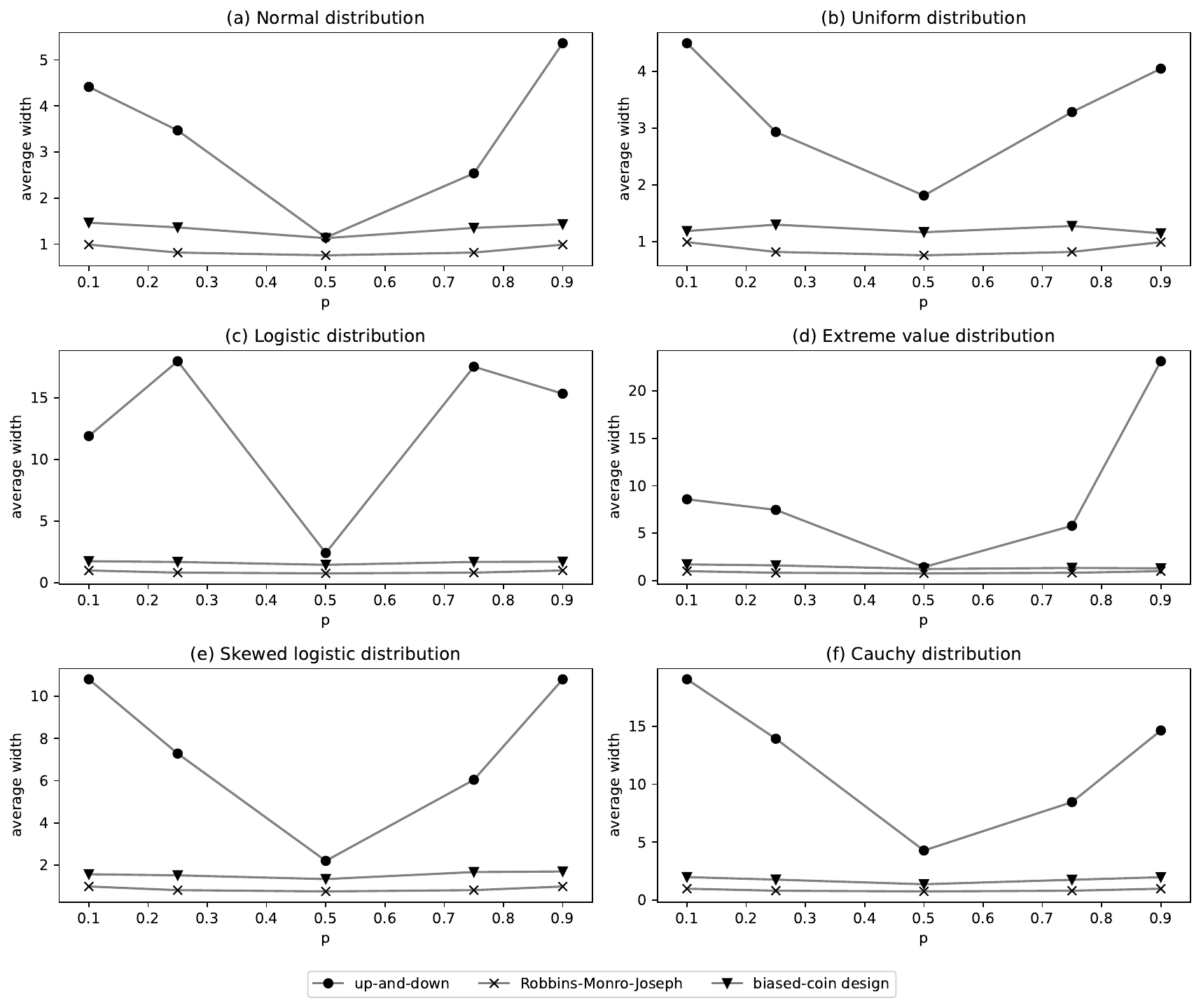}
    \end{center}
    \caption{Average width of confidence intervals from the simulation study after $n=30$ trials.}
    \label{fig:width}
\end{figure}

We now move to confidence intervals. First we compare the average width of the confidence intervals yielded by the three approaches, which are plotted in Figure~\ref{fig:width} for $n=30$. Again, the largest discrepancy is between the up-and-down procedure and the rest, but here, the RMJ procedure consistently outperforms the BCD, albeit by quite a small margin. As for the MSEs, the results for $n=100$ are very similar and are included in the Appendix.

\begin{figure}
    \begin{center}
        \includegraphics[scale=.52]{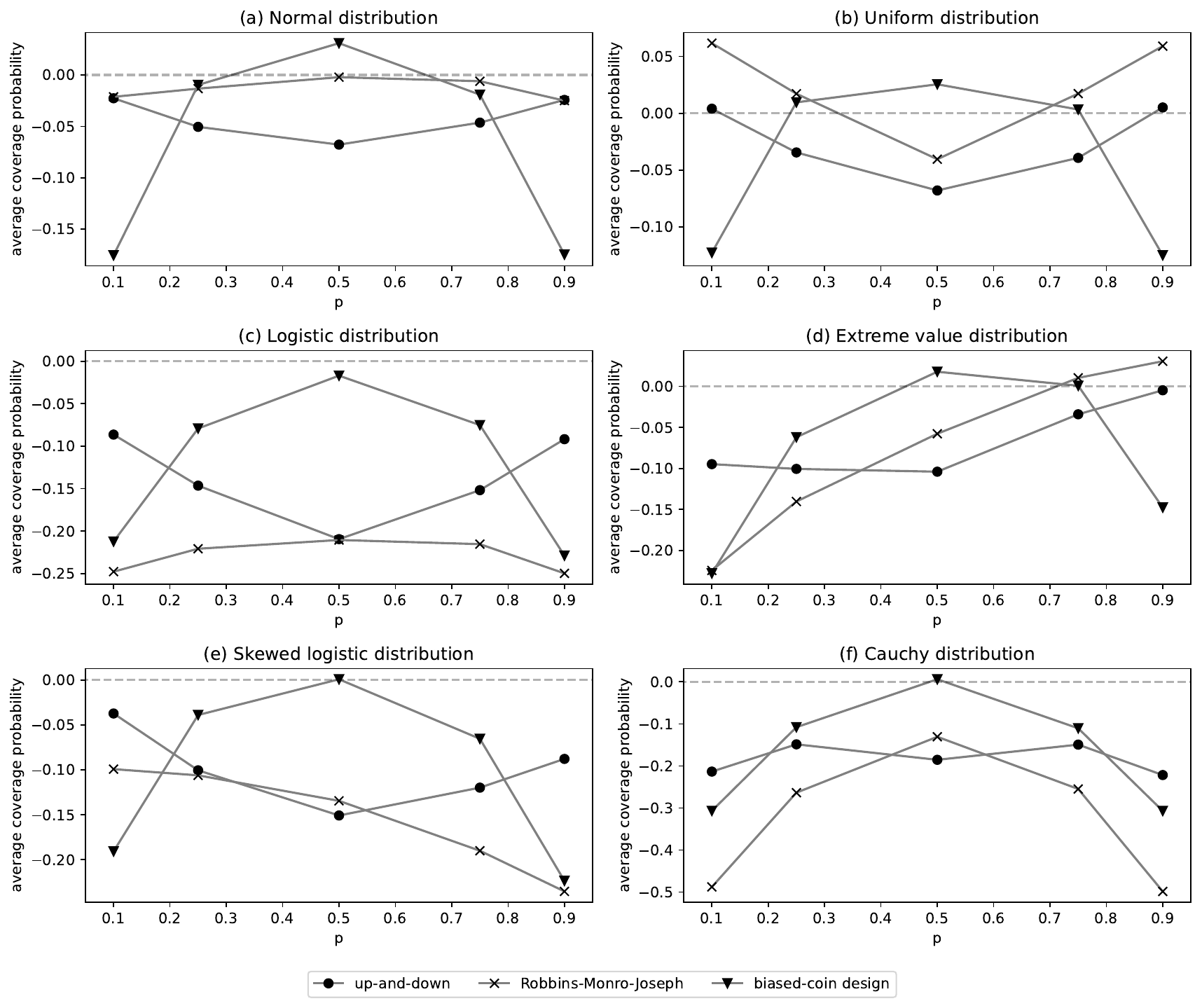}
    \end{center}
    \caption{Average coverage probability of 90\% confidence intervals from the simulation study after $n=30$ trials.}
    \label{fig:covprob1}
\end{figure}

\begin{figure}
    \begin{center}
        \includegraphics[scale=.52]{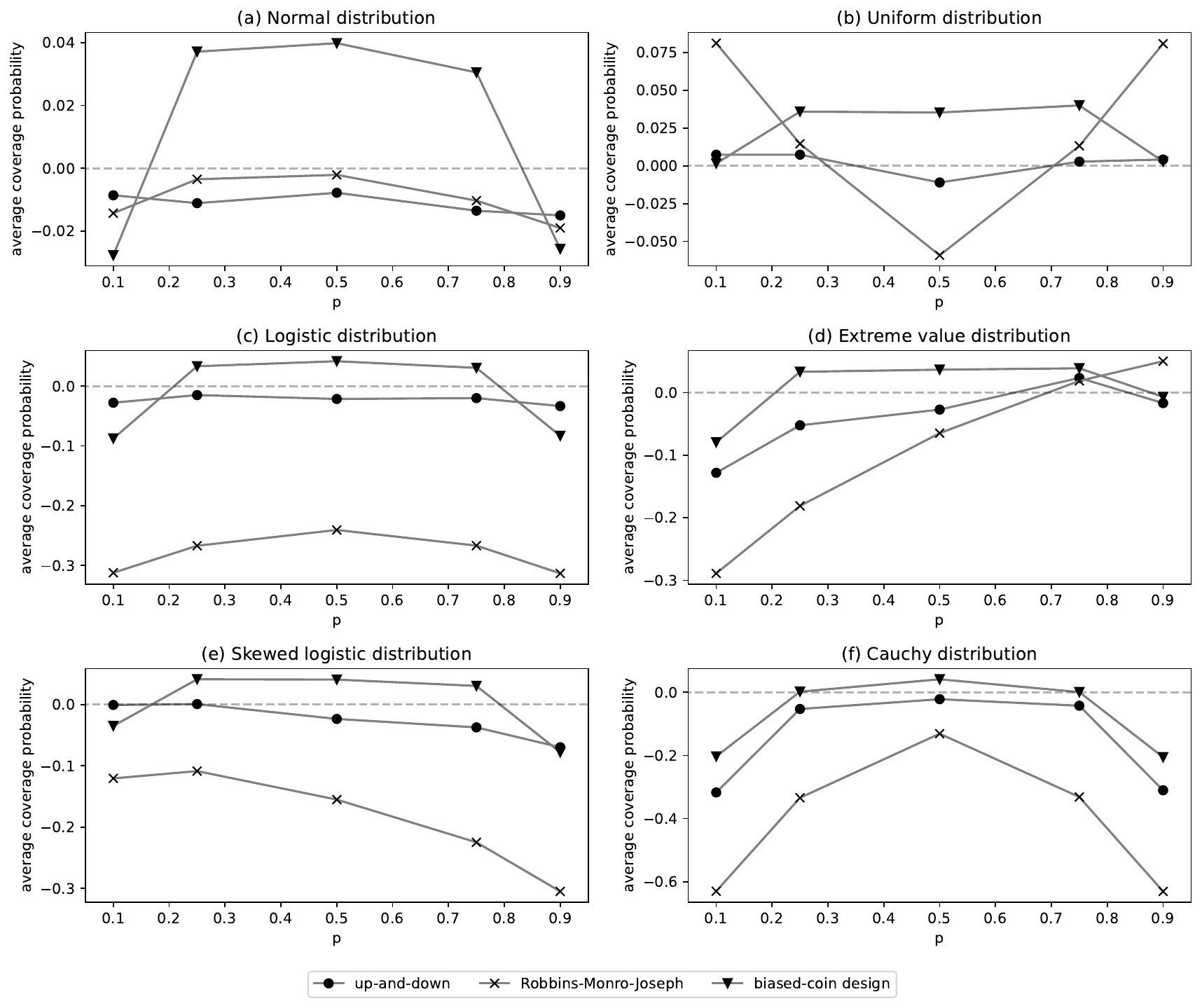}
    \end{center}
    \caption{Average coverage probability of 90\% confidence intervals from the simulation study after $n=100$ trials.}
    \label{fig:covprob2}
\end{figure}

\begin{figure}
    \begin{center}
        \includegraphics[scale=.52]{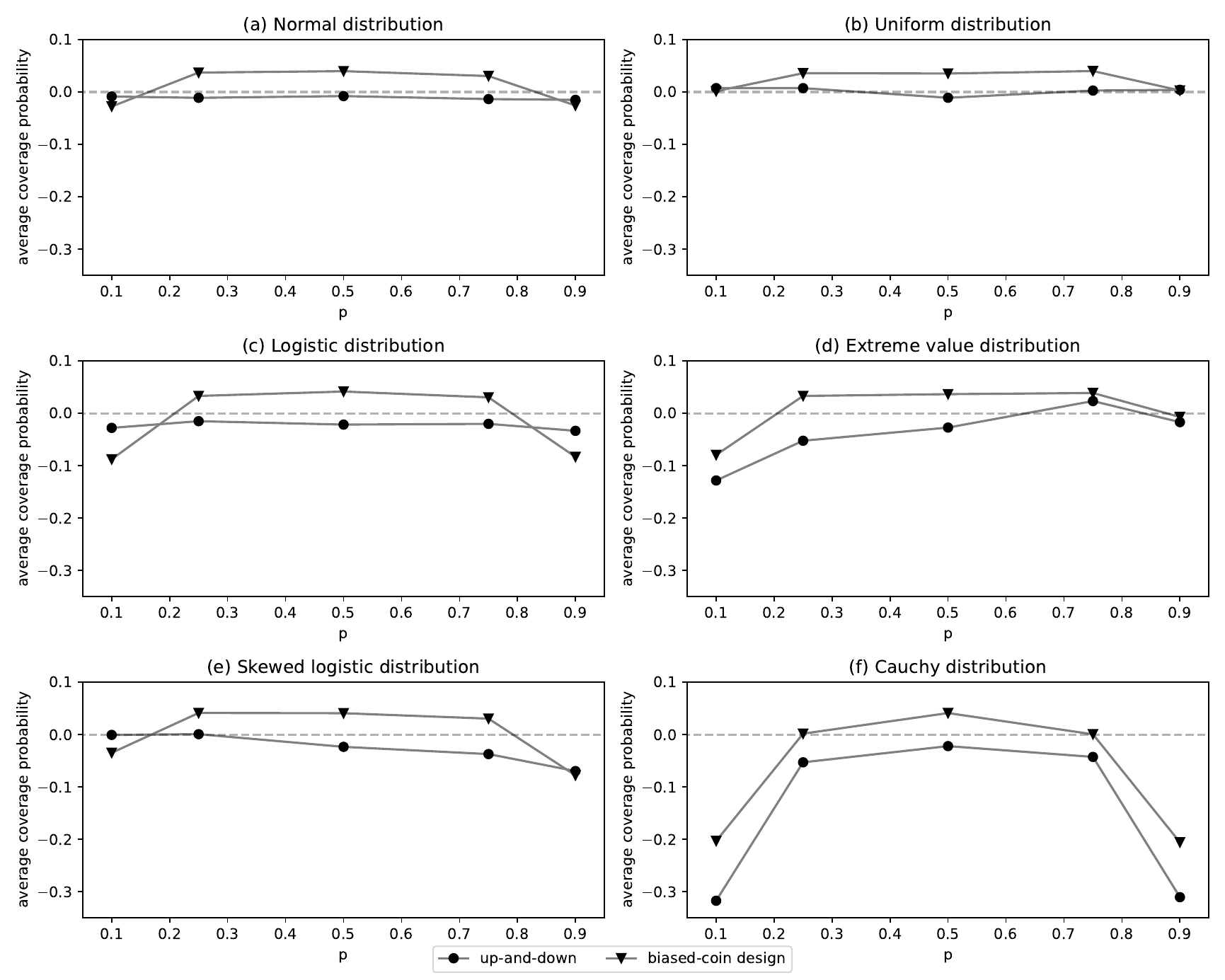}
    \end{center}
    \caption{Average coverage probability of 90\% confidence intervals from the simulation study after $n=100$ trials, showing up-and-down and BCD only.}
    \label{fig:covprob3}
\end{figure}

Comparing the widths of confidence intervals does not make much sense unless we also compare their empirical coverage probabilities, i.e.~what proportions of the confidence intervals generated in the simulations actually covered the target quantile. Here the picture becomes more nuanced. Figure~\ref{fig:covprob1} shows the coverage probabilities obtained after $n=30$ trials. At first glance, it seems like all three methods perform equally poorly. The only apparent consistent pattern is that the BCD does very well for $\xi_{50}$, but then undershoots for more extreme quantiles. However, if we also look at the results after $n=100$ trials, shown in Figure~\ref{fig:covprob2}, a clearer pattern emerges. Here we see that the up-and-down and BCD approaches consistently outperform the RMJ procedure, except in the normal model, where $\dot{M}(0) = \phi(\Phi^{-1}(p))$ is correctly specified. To make the comparison between the up-and-down and BCD approaches easier, see Figure~\ref{fig:covprob3}, where we have fixed the scaling on the $y$ axis and removed the RMJ results. Now we see that in fact, the overall best performing confidence intervals are those constructed from Fieller's theorem for the classic up-and-down design. The coverage probabilities obtained by this approach are particularly satisfactory for the first three model choices. However, when the model diverges far from normality (i.e.~when the probit model is severely misspecified), such as for the extreme value and Cauchy distributions, the performance of Fieller's theorem is more unstable than that of the BCD. This is especially true for extreme quantiles such as $p=0.1$ and $p=0.9$.

Based on these simulations, we see that both the RMJ and the BCD provide better point estimates for $\xi_{100p}$ than the up-and-down design equipped with maximum likelihood estimation, even when the latter model is correctly specified. We also see neither method yields satisfactory coverage probabilities after only $n=30$ observations, and so we recommend a larger sample size like $n=100$ if the confidence interval is to be trusted. If the researcher has reason to believe that the sensitivity distribution is close to normal, then the best approach is to use CIR for point estimation and Fieller's theorem for confidence intervals. If the researcher is agnostic about the shape of the distribution, then the BCD is the best performing method. 

\subsection{PETN revisited}
Let us now return to the problem of estimating the friction sensitivity of PETN, which we first did using the UN manual in Section~\ref{sec:PETN1}. Based on the results of the simulation study from the previous section, we employed the BCD targeting the quantile $F_{10}$. We chose $x_1 = \SI{80}{J}$ as starting value, based on the PETN reference from \citet{novik2025characteristics}. We first did one test with a step size of $d = \SI{20}{J}$ with $n=30$ trials and then a second test with a smaller step size of $d=\SI{8}{J}$ with $n=50$ trials. In the second test, we observed a negative reaction at \SI{80}{J}, requiring escalation to \SI{88}{J}. However, since the load \SI{88}{J} is not available in Table~\ref{tab:frictional-loads}, we used \SI{96}{J} instead (one could also have used \SI{84}{J}). It is important to note that CIR works for uneven step sizes. The data from both tests are given in Tables~\ref{tab:BCD-PETN1} and~\ref{tab:BCD-PETN2}.

\begin{table}
    \centering
    \caption{Results of friction sensitivity testing of PETN using the BCD with $n=30$ and $d=20$.}\label{tab:BCD-PETN1}
    \begin{tabular}{lccc}
        \hline
        Load (N) & 40 & 60 & 80 \\
        \# Reactions & 0 & 2 & 2 \\
        \# Trials & 9 & 19 & 2 \\
        \hline
    \end{tabular}
\end{table}

\begin{table}
    \centering
    \caption{Results of friction sensitivity testing of PETN using the BCD with $n=50$ and $d=8$ (except for one step, see details in the text).}\label{tab:BCD-PETN2}
    \begin{tabular}{lcccccccc}
        \hline
        Load (N) & 32 & 40 & 48 & 56 & 64 & 72 & 80 & 96 \\
        \# Reactions & 0 & 1 & 1 & 1 & 2 & 2 & 4 & 1 \\
        \# Trials & 3 & 8& 3 & 5 & 8 & 14 & 6 & 3 \\
        \hline
    \end{tabular}
\end{table}

The CIR point estimates for the two datasets are \SI{58.95}{J} and \SI{38.17}{J}, respectively. These are quite different results, which reflects both that $F_{10}$ is more difficult to estimate than $F_{50}$, and that the step size \SI{20}{J} is probably too coarse for PETN. However, there is still wide overlap between the two 90\% confidence intervals obtained, which are [\SI{23.47}{J}, \SI{61.87}{J}] and [\SI{18.83}{J}, \SI{61.79}{J}], respectively. Either analysis provides significantly more information about the friction sensitivity of PETN than the results from Section~\ref{sec:PETN1}.

\section{Conclusion}
We have thoroughly demonstrated that the limiting stimulus is an ill-founded notion of sensitivity. This means that the sensitivity tests in the UN manual depend heavily on arbitrary experimental parameters, directly opposing the proclaimed purpose of the manual. We have shown that the continued use of limiting stimuli causes confusion in energetic materials research and obstructs the statistical analysis of the sensitivity of explosives. We have demonstrated that substantially better alternatives are available in the sensitivity testing and pharmacology literature, with the biased coin design performing the best overall in our simulations. By switching from limiting stimulus estimation to approaches like the biased coin design targeting a specific quantile, researchers will be estimating a well-founded and interpretable notion of sensitivity, with an accompanying confidence interval to address the uncertainty of the estimate.

% Bibliography
\bibliographystyle{plainnat-modified}
\bibliography{bibliography}

\newpage
\appendix
\section*{Appendix}
\subsection*{Additional figures}

\begin{figure}[H]
    \begin{center}
        \includegraphics[scale=.52]{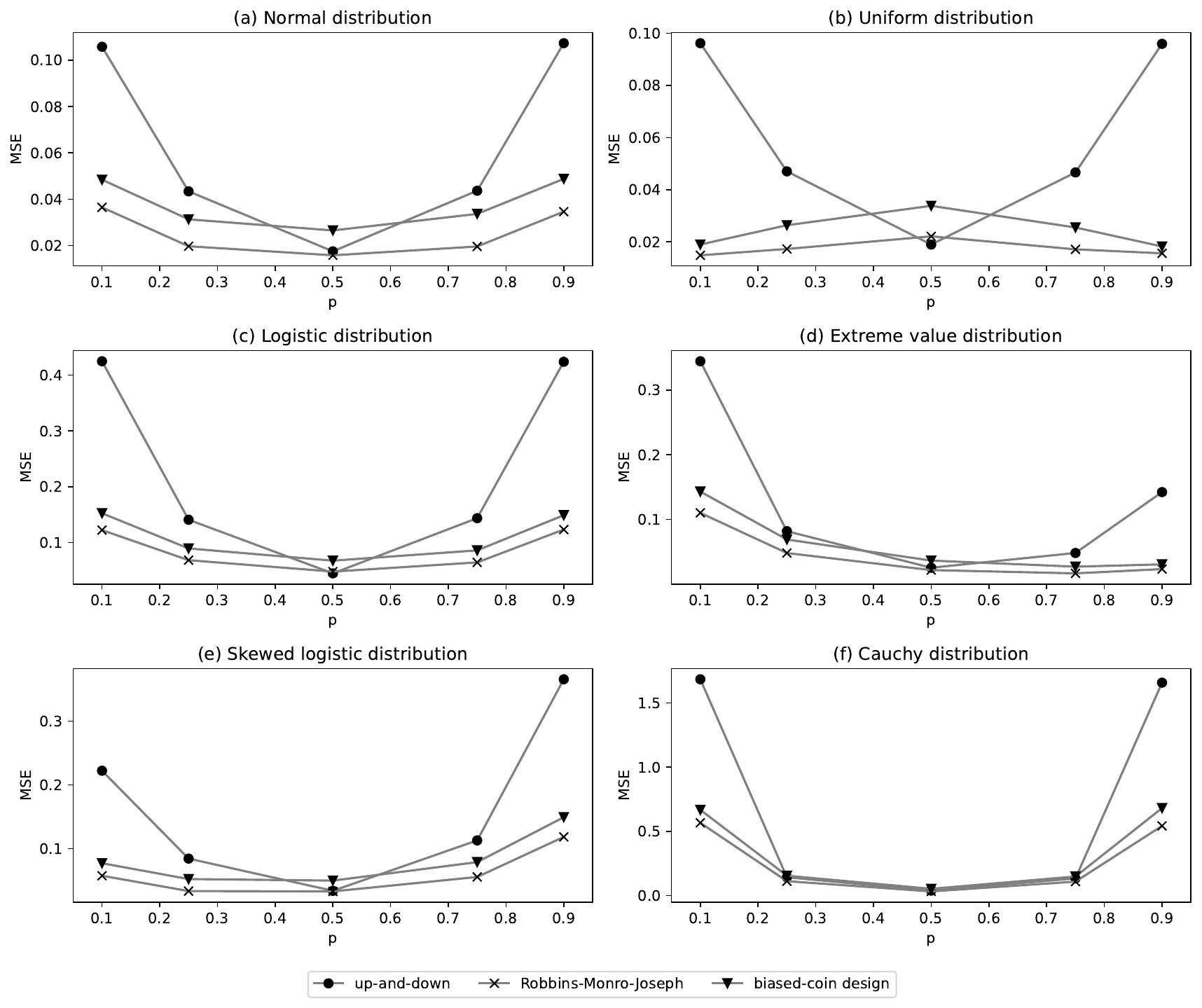}
    \end{center}
    \caption{Mean squared errors (MSEs) from the simulation study after $n=100$ trials.}
    \label{fig:MSE2}
\end{figure}

\begin{figure}[H]
    \begin{center}
        \includegraphics[scale=.52]{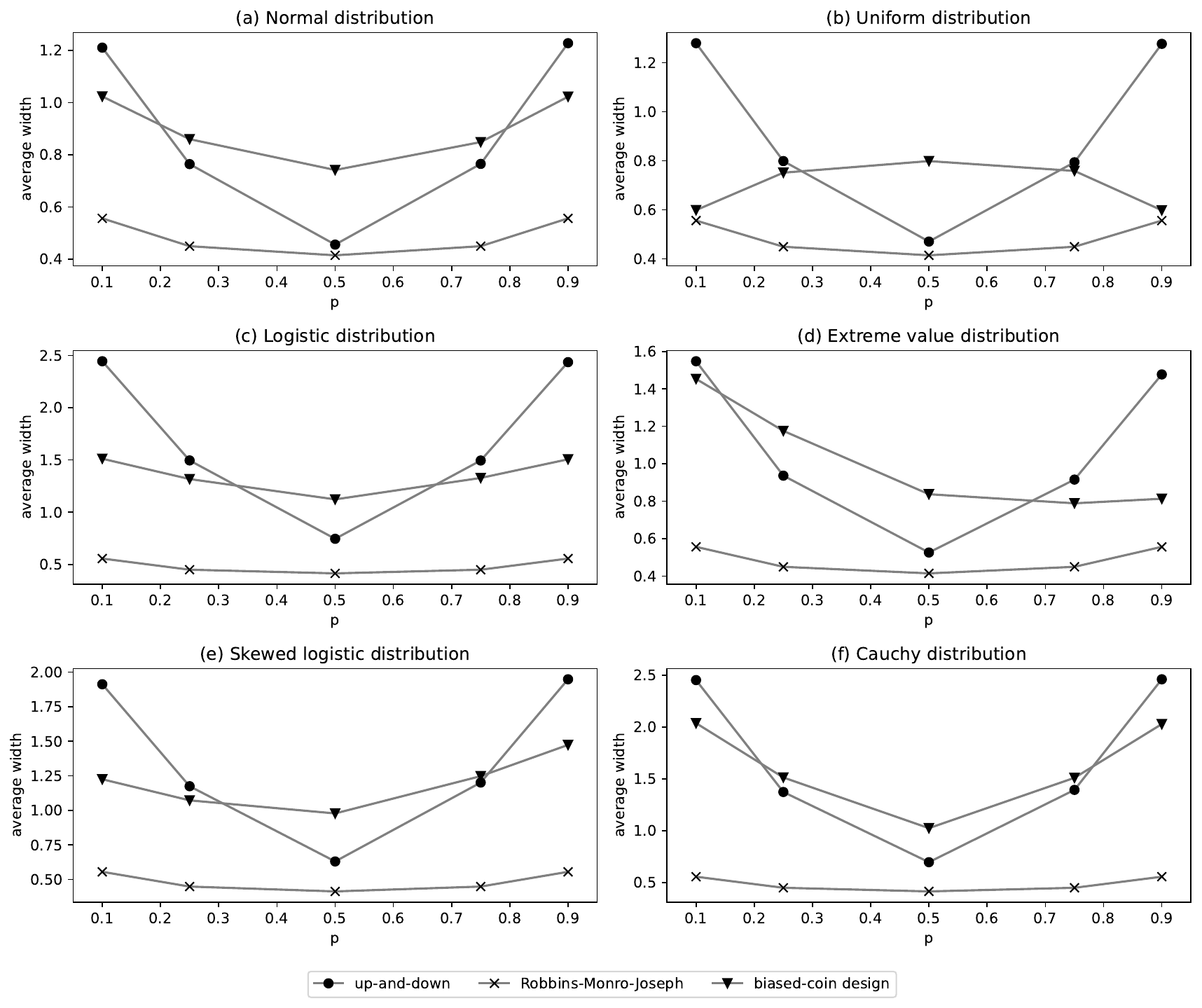}
    \end{center}
    \caption{Average width of confidence intervals from the simulation study after $n=100$ trials.}
    \label{fig:width2}
\end{figure}

\end{document}